# A Novel Multi-Criteria Local Latin Hypercube Refinement System for Commutation Angle Improvement in IPMSMs

Pedram Asef, *Senior Member, IEEE*, Mouloud Denai, Johannes J. H. Paulides, *Senior Member, IEEE,* Bruno Ricardo Marques, and Andrew Lapthorn *Senior Member, IEEE*

*Abstract*—The commutation angle, γ, of an interior permanent magnet synchronous motor's (IPMSM) vector diagram, plays an important role in compensating the back electromotive force (back-EMF); both under phase current variations and an extended speed range, is required by the application. This commutation angle is defined as the angle between the fundamental of the motor phase current and the fundamental of the back-EMF. It can be utilised to provide a compensating effect in IPMSMs. This is due to the reluctance torque component being dependent on the commutation angle of the phase current even before entering the extended speed range. A real-time maximum torque per current and voltage strategy is demonstrated to find the trajectory and optimum commutation angles, γ, where the level of accuracy depends on the application and available computational speed. A magnet volume reduction using a novel multi-criteria local Latin hypercube refinement (MLHR) sampling system is also presented to improve the optimisation process. The proposed new technique minimises the magnet mass to motor torque density whilst maintaining a similar phase current level. A mapping of γ allows the determination of the optimum angles, as shown in this paper. The 3rd generation Toyota Prius IPMSM is considered as the reference motor, where the rotor configuration is altered to allow for an individual assessment.

## I. NOMENCLATURE

| | |
|---|---|
| $v_d, v_q$ | *dq*-axis voltages transposed fundamental abc voltages |
| $i_d, i_q$ | *dq*-axis current transposed fundamental abc voltages |
| $L_d, L_q$ | *dq*-axis inductances transposed fundamental abc voltages |
| $\lambda_m$ | Flux linkage produced by permanent magnets |
| $R_s$ | Stator winding phase resistance |
| $i_s$ | Stator instantaneous current |
| $\omega_e$ | Electrical angular velocity |
| $\gamma$ | Commutation angle |
| $i_{ph}$ | Fundamental of the phase current |
| $e_{ph}$ | Fundamental electromotive force |
| $T_e$ | Electromagnetic torque |
| $I_m$ | Maximum current limited by the inverter |
| $V_m$ | Maximum voltage limited by the inverter |
| $R_w$ | Vehicle wheel radius |
| $\eta_{trans}$ | Transmission efficiency |
| $G_r$ | Transmission gear ratio |
| $F_t$ | Total longitudinal (or resistive) force |
| $a_{x,max}$ | Maximum longitudinal acceleration |
| $F_{x,F/R}$ | Front or rear longitudinal tire force in traction |
| $F_{z,F/R}$ | Vertical load on the front or rear axle |
| $g$ | Gravitational acceleration |
| $m_{app}$ | Apparent mass of the rotating components of powertrain |
| $\vartheta_{max}$ | Maximum longitudinal road gradient |
| $T_{m,F/R}$ | Front or rear mechanical torque in traction |
| $m_0, m_1$ | Vehicle gross mass and payload mass |
| $C_r$ | Rolling resistance coefficient |
| $T_{m,max}$ | Maximum mechanical torque in traction |
| $T_{ref}$ | Reference torque |
| $p$ | Number of pole pairs |
| $\phi_p$ | Space-filling criterion driving LHS method |
| $d_{ij}$ | Distance between two samples in LHS method |
| TPCA | Maximum torque per commutation angle ratio |

This paragraph of the first footnote will contain the date on which you submitted your paper for review, which is populated by IEEE. It is IEEE style to display support information, including sponsor and financial support acknowledgment, here and not in an acknowledgment section at the end of the article. *(Corresponding author: P.Asef)*

P.Asef is with the Dept. of Electronic and Electrical Engineering, University of Bath, BA2 7AY UK (e-mail: pa696@bath.ac.uk).

M. Denai. is with Dept. of Engineering and Technology, University of Hertfordshire, AL10 9AB UK (e-mail: m.denai@herts.ac.uk).

J. J. H. Paulides is with Advanced Electromagnetics Group, Waalwijk, The Netherlands (e-mail: johan@ae-grp.nl).

B. R. Marques is with AE UK Ltd part of AE-Group, Nottingham, UK (e-mail: bruno.marques@ae-grp.co.uk).

A. Lapthorn is with Dept. of Electrical and Computer Engineering, University of Canterbury, Christchurch, New Zealand (e-mail: andrew.lapthorn@canterbury.ac.nz).

## II. INTRODUCTION

a) *Motivation and Background*

In IPMSMs, the rotor design is one of the most complex parts; it can be noted that the rotor reluctance circuit with its specific permanent magnet (PM) arrangement strongly influences the overall electromagnetic contribution of the powertrain. The complete system assessment is outside the scope of this paper. The permanent magnets' embedded or buried nature provides mechanical rigidity, despite the significant centrifugal force at high speed, and produces a hybrid torque via a combination of magnet and reluctance contributions, resulting in increased efficiencies over the complete torque-speed characteristic [1].

b) *Relevant Literature*

Numerous complex magnet arrangements have been proposed for achieving specific application-related performance characteristics [2-3]. For example, in [4], Guo and Parsa investigated the influence of straight, V- and U-shaped magnet arrangements on the torque characteristics of IPMSMs. In [5], Dlala et al. considered the efficiency map of a PM machine with a V-shaped magnet layout and evaluated the achievable magnet size reduction. Ishikawa et al. [6] studied the dependency of the torque ripple on the shape and location of the permanent magnets. They implemented an optimisation method using the response surface methodology to improve the vibrational aspects of the machines. Similarly, Jung et al. [7] successfully reduced the noise and vibrations of a specific machine prototype by optimising the geometry of the magnet flux barriers.

Technically, IPMSMs, with a rotor saliency of $L_q L_d^{-1} > 1$, can offer appealing performance characteristics, such as providing flexibility for adopting a variety of rotor geometries (spoked or embedded/buried magnets) as alternatives to non-salient surface mounted synchronous machines. Buried magnets inside the rotor core provide the basis for a more mechanically robust rotor construction capable of higher speeds since the permanent magnets are physically contained and protected inside the rotor lamination. However, in electromagnetic terms, introducing such laminated steel pole pieces above and next to the permanent magnets fundamentally alters the machine's reluctance circuit. Altering the rotor reluctance flux tube paths introduces additional rotor saliency that can reduce the PM volume in the IPMSMs while maintaining the torque-speed operating range.

The complex nonlinear nature of an electric motor with, in this paper considered, rotor saliency requires an optimised maximum torque per current and voltage MTPA/MTPV strategies to enhance the electric drive efficiency as this directly impacts range, acceleration, and deceleration performance. It needs noting that stator saliency is not altered nor optimized. Among all published approaches [8-12], (i) mathematical-based methods [8-9], (ii) look-up table methods [10], (iii) online search methods using optimisation methods [11], and (iv) signal injection methods [12] are the most popular strategies to enhance the electric drive effectiveness. Depending on the complexity, each approach offers advantages in terms of computational time and accuracy of the selected strategy. Most analytical-based techniques like (i) and (ii) neglect the partial derivatives of the IPMSMs subjected to the commutation angle. Using such strategies limits the MTPA/MTPV ability to compute the operation points more accurately, and thus, deviations from the optimum torque or efficiency are likely. Recently, approaches (iii) and (iv) have become more popular due to their ability to increase electric drive efficiencies or effectiveness depending on the selected optimization criteria.

c) *Contributions and Organisation*

This paper analyses the impacts of rotor configurations using the fundamental of the stator continuous current and back electromotive force (back-EMF) vector diagrams. Various interior permanent magnet (IPM) rotor configurations are assessed for a pure electric vehicle (EV) with an all-wheel-drive configuration. The commutation angle maps for every IPM rotor configuration are investigated to introduce a factor that represents the highest maximum torque per commutation angle. Following the selection of the "best" IPMSM, a finite element analysis (FEA) is performed to verify the results with the EV drivability indexes at the vehicle level including also higher order harmonics. The analysis determines whether the proposed analytical factor is valid to make initial predictions on which IPM rotor configuration can improve vehicle acceleration and longitudinal road gradient capabilities. The vehicle dynamic parameters and configuration are given for the studied IPMSMs [13]. The paper aims to improve the maximum Torque Per Commutation Angles (TPCA). To satisfy this aim, the following objectives of the work are required: (i) to develop a novel multi-criteria local Latin hypercube refinement (MLHR) sampling system to improve the optimisation process for sizing the magnets (ii) to map and evaluate the energy efficiencies and commutation angles, γ, variations and study its impact on the efficiency that is mapped on the torque-speed profile using a 2-D FEA of various embedded/buried IPM rotor configurations; (iii) each rotor configuration's capacity is analysed, subject to its drivability metrics such as vehicle acceleration and longitudinal road gradient. The EV considered in this study is a four-wheel-drive electric passenger car with two identical onboard IPMSMs, one per axle. Each IPMSM is coupled to the two wheels of the axle through a single-speed mechanical transmission with an open differential, half-shafts, and constant velocity joints; and (iv) introducing a new factor, computed by MLHR algorithm, reporting the maximum TPCA for all three main regions, low speed, accelerating, and high-speed operations. The 3rd generation Toyota Prius (T-Prius) IPMSM is considered as the reference motor for this study. All the studied IPMSMs are assessed under both urban and motorway drive cycles to evaluate the vehicle's drivability performance during acceleration and deceleration.

III. ANALYTICAL AND NUMERICAL MODELLING

a) *IPMSM Mathematics Considering Commutation Angle*

Considering the fundamental waveforms, mathematical models of the IPMSMs are well formulated [14-15], in which the *d-q* axis voltages under the steady-state condition are:

$$\begin{bmatrix} v_d \\ v_q \end{bmatrix} = \begin{bmatrix} R_s & -\omega_e L_q \\ \omega_e L_q & R_s \end{bmatrix} \begin{bmatrix} i_d \\ i_q \end{bmatrix} + \begin{bmatrix} 0 \\ \omega_e \lambda_m \end{bmatrix} \quad (1)$$

The commutation angle for *dq*-axis currents:

$$\begin{cases} i_d = -i_s \sin\gamma \\ i_q = i_s \cos\gamma \end{cases} \quad (2)$$

where the fundamental electromotive force, $e_{ph}$, ($\gamma = (i_{ph}, e_{ph})$).

The electromagnetic, $T_e$, can be calculated using [14-15]:

$$T_e = \frac{3}{2} p [\lambda_m + (L_d - L_q)i_d] i_q \quad (3)$$

with consideration of the partial derivative of commutation angle, γ:

$$\frac{\partial T_e}{\partial \gamma} = \frac{3p}{2} \Big[ -\lambda_m i_s \sin\gamma + \frac{\partial \lambda_m}{\partial \gamma} i_s \cos\gamma - L_d i_s^2 \cos2\gamma + L_q i_s^2 \cos2\gamma - \frac{\partial L_d}{\partial \gamma} \frac{i_s^2}{2} \sin2\gamma + \frac{\partial L_q}{\partial \gamma} \frac{i_s^2}{2} \sin2\gamma \Big] \quad (4)$$

A real-time Maximum Torque Per Current (MTPA) and Maximum Torque Per Voltage (MTPV) control strategy is employed to satisfy the EV driver demands. Such a strategy is reported in [16], where the authors used the rate of change of the parameters of the IPMSM's given by Eq. (1)-(3) to optimise the working criteria of both MTPA and MTPV strategies using a Lagrange multiplier. This multi-objective problem is defined as:

$$min\, f(i_d, i_q) = f_1 + f_2 \quad (5)$$

The first objective is the MTPA which can be computed as an optimisation problem using [16]:

$$min\, f_1(i_d, i_q) = (-i_s \sin\gamma)^2 + (i_s \cos\gamma)^2 \quad (6)$$

subject to the following constraint:

$$s.t. \begin{cases} T_{ref} - \frac{3}{2}p\left[\lambda_m + \left(L_d(i_d, i_q) - L_q(i_d, i_q)\right)i_d\right] = 0 \\ i_d^2 + i_q^2 \leq I_m^2 \end{cases} \quad (7)$$

The second objective is to satisfy the MTPV trajectory based on the following cost function [16]:

$$min\, f_2(i_d, i_q) = v_d^2 + v_q^2 \quad (8)$$

subjected to the following constraint:

$$s.t. \begin{cases} T_{ref} - \frac{3}{2}p\lambda_q \left[\frac{\lambda_m}{L_q(i_d,i_q)} + \lambda_d \left(\frac{1}{L_q(i_d,i_q)} - \frac{1}{L_q(i_d,i_q)}\right)\right] = 0 \\ \sqrt{v_d^2 + v_q^2} \leq V_m \end{cases} \quad (9)$$

The objective function minimisation is essentially a torque maximisation at a certain speed within the constant torque region (below base speed) once the torque demand is satisfied by considering the limitations of phase or line-to-line voltages and currents or by altering the commutation angle. By increasing the speed demand (above base speed), the torque demand could be achieved by decreasing the phase current (recall that in a star connected winding phase and line current are the same) below the maximum allowed value or by altering the commutation angle.

*b) Latin Hypercube Sampling (LHS) for Magnet Sizing*

In the numerical FEA modelling, the stator topology and specification are fixed, for the fairness of the study, based on a 3rd generation Toyota Prius motor and reported in Table I for the fairness of the study. The rotor configurations are only changed via different IPM magnet arrangements, and these initial parameters are summarised in Table II. The main dimensional parameters in Tables I and II are selected based on the original 3rd generation Toyota Prius motor. Some parameters are defined within a range; all design variables are constrained to the reference motor's maximum magnet volume, such as $W_{w1}$, $T_m$, $L_m$, $W_m$, $W_g$, $\alpha_{m1}$, $\alpha_{m2}$, and $\alpha_{m3}$. For the modified V2, these PM-related parameters, for example, create the controllable input vector $X^0 = (W_m, T_m, L_m, W_{w1}, W_g, \alpha_{m1}, \alpha_{m2}, \alpha_{m3})$ with eight variables, and consideration of noise per variable $N_x = (N_{W_m}, N_{T_m}, N_{L_m}, N_{W_m}, N_{W_g}, N_{\alpha_{m1}}, N_{\alpha_{m2}}, N_{\alpha_{m3}})$. Hence, the output is a function of $Y = f(X, N_x, i_d, i_q)$. The space-filling is done uniformly using a criterion, as proposed in [17-18], to initialise the samples (which is set to $n = 100$) using random uniform distribution for the LHS probabilistic algorithm:

$$\phi_p = \left(\sum_{i=1}^{n-1}\sum_{j=i+1}^{n=100} d_{ij}^{-p}\right)^{1/p} \quad (10)$$

where $n$ is the total number of samples, $d_{ij}$ is the measured distance between two random samples, and $p$ is a coefficient suggested to be set $p = 50$ in [18-19]. The LHS algorithm, as its pseudo coding is shown by Algorithm 1, attempts to maximise $d_{ij}$ to satisfy the space-filling process, which means minimising the criterion $\phi_p$. The distance between samples is:

$$d_{ij} = \left(\sum_{k=1}^{n_v} |x_{ik} - x_{jk}|^t\right)^{1/t} \quad (11)$$

TABLE I MAIN STATOR PARAMETERS

| Parameter | Description | Unit | Value |
|---|---|---|---|
| $D_{si}$ | Inner stator diameter | mm | 161.9 |
| $D_{so}$ | Outer stator diameter | mm | 264 |
| $H_s$ | Slot height | mm | 30.9 |
| $W_s$ | Slot width | mm | 6.69 |
| $H_1$ | Intermediary height of the slot | mm | 0.27 |
| $L$ | Axial length | mm | 50 |
| $L_g$ | Airgap length | mm | 0.75 |
| $N_s$ | Number of slots | - | 48 |
| $R$ | Slot bottom radius | mm | 3.345 |
| $V$ | Undercut angle of stator tooth tip | deg | 20.298 |
| $W_{s2}$ | Top slot width | mm | 6.69 |
| $W_{s1}$ | Bottom slot width | mm | 3.34 |
| $W_{T1}$ | Tooth width, upper part of slot | mm | 7.45 |
| $W_{T2}$ | Tooth width, bottom part of slot | mm | 7.502 |
| $W_o$ | Width of slot opening | mm | 1.88 |
| $H_o$ | Height of slot opening | mm | 1.22 |

TABLE II MAIN ROTOR PARAMETERS WITH THE PM SIZE RANGES

| Parameter | Description | Unit | T.Prius | Rotor 2 & 3 |
|---|---|---|---|---|
| $D_{ro}$ | Outer rotor diameter | mm | 160.4 | 160.4 |
| $D_{ri}$ | Inner rotor diameter | mm | 110 | 110 |
| $p$ | Number of poles | - | 8 | 8 |
| $W_{w1}$ | Window width | mm | 0.7 | 0.2-1.2 |
| $T_m$ | Magnet thickness | mm | 7.2 | 2-7.16 |
| $T_1$ | Thickness of rotor yoke under the magnet | mm | 8.14 | 8.14 |
| $L_m$ | Axial length of magnet | mm | 50 | 45-50 |
| $V_{pm}$ | Magnet volume | cm³ | 6.401 | <6.4 |
| $V_p$ | Angular pitch | deg | 45 | 45 |
| $W_m$ | Magnet width | mm | 17.88 | 14-25 |

**Algorithm 1**: LHS Pseudo into NSGA-II for Magnet Sizing

1: Range the variation of each input $x$: i.e. $x_i = (x_{i_{min}}, x_{i_{max}}) \in \mathbb{R}$
2: Variables space: $\wp = x_1 \times x_2 \times x_3 \times x_4 \times x_5 \times x_6 \times x_7 \times x_8 \subset \mathbb{R}^3$
3: Create matrix $S$ with size $(nP \times jj)$;
4:     If $n = 2n$ **use** $ii = \frac{nP}{2}$
5:     *Else*, **set** $ii = \frac{nP-1}{2}$ and update $A\left(\frac{nP+1}{2}, j\right) = \frac{nP+1}{2}$   *end if*
6: Set $n$ and $n_v$ (sample distribution interval set [0,1]
7: *For* $i = 1$ to $ii$ & $j = 1$ to $jj$
8:     Initialise $X \in X_0, X_{opt} = X_{i+1}, X_{i-1} = X_{opt}$
9:     Generate samples by solving Eq. 10 to select the best $\phi_p$
10:      If $n < 0.5$, set $A(i,j) = \varepsilon_j(i)$,
11:       *otherwise* set $A(i,j) = nP + 1 - \varepsilon_j(i)$ *end if*
12:      If $min\, f(i_d, i_q) = f_1 + f_2$   #optimisation cost function
13:        *s.t.* $V_{pm_{new}} < V_{pm_{old}}$
14:        $A_{\eta_{opt}} \geq A_{\eta_0}$
15:        $TPV_{opt} > \frac{T_{e_i}}{V_{pm_i}} > TPV_0$
16:      *Else* re-simulate $f$ based on new set of $X$, *end*
17: Store all new sets of $X_{opt}$, considering IPMSM performance where $n_v$ is the dimension of samples $n$, $x_{ik}$ is the $k$th component of $i$ sample, $x_{jk}$ is the $k$th component of $j$ sample, and $t$ coefficient is set to 1, based on [18-19]. The LHS sampling process for a global optimisation, considering Eq. 10 and FEA outputs (e.g. torque, power, premium efficiency region area) from the IPMSMs, is summarised in Table III.

c) *Proposed MLHR System for Magnet Sizing*

The reason behind developing the MLHR system is improving sample efficiency. The developed MLHR software can do multi-criteria design optimisation using a DOE/R-based sensitivity analysis, 2-D FEA, and an improved sample efficiency for a heuristic-based optimisation method like NSGA-II. The machine's parameters are defined within a Python code and the FEA is employed to carry out the electromagnetic analysis. The sensitivity and optimisation methods are also incorporated in the MLHR software. Magnet sizing, like other electrical machine optimisation problems, is a high dimensionality problem, that requires an extensive volume of samples. Compared to the traditional LHS method, the developed MLHR can work with affordable sample numbers by increasing the sample efficiency. MLHR uses an agnostic sequential sampling scenario to discover a surrogate-focused optimum point for dealing with multi-criteria and multi-objective robust optimisation problems in this algorithmic system. To do this, both Gaussian and support vector regression (SVR) models are utilised as surrogate models. In such a model, the algorithm estimates the relationship $\tilde{f}(x) = \tilde{y}$ subjected to the selected variables $X^0 \in \mathbb{R}^{m \times n}$ and its related system $y^0 \in \mathbb{R}^m$, where $m$ is the number of samples and $n$ is the number of variables. Next, a kernel model of $k(x_1, x_2): \mathbb{R}^{2n} \to \mathbb{R}$ is defined only to measure the distance between two samples $x_1$ and $x_2$. Based on this surrogate model, the estimation Gaussian function is:

$$\tilde{f}(x_{pred}) = \mu_Y(x) = k(C + \sigma_\epsilon^2 I_m)^{-1} y^0$$
$$k = [k_1, \ldots, k_m] \quad (12)$$
$$k_i = k(x_{pred}, x_i^0; \theta^h, \sigma^2)$$
$$C_{:,i} = C_{:,i}^T = k(X_{i,:}^0, X^0; \theta^h, \sigma^2)$$

where $x_{pred}$ is the predicted variable sample, $\mu_Y$ is a Gaussian parameter with a defined mean, $\sigma_\epsilon$ is the measurement noise that occurred in the samples $x_i^0 = X_{i,:}^0$ and output data $y^0$. $C_{:,i}$ and $C_{i,:}$ are the $i$th row and column of the covariance matrix $C$. $\theta^h$ is indicating the vector of kernel parameters that is $\theta^h \in \mathbb{R}^n$, $\sigma^2$ is the constant process variance. Based on [20-21], the maximum likelihood prediction is investigated, and the log-likelihood of the parameters given in (12) are maximised, as given:

$$\mathcal{L}(\theta^h, \sigma, \sigma_\epsilon) = -\frac{m}{2}\log(2\pi) - \frac{\log(|C+\sigma_\epsilon^2 I_m|)}{2} - \frac{y^{0,T}(C+\sigma_\epsilon^2 I_m)^{-1} y^0}{2} \quad (13)$$

Considering the SVR prediction function:

$$\tilde{f}(x) = \sum_i^m c_i k(x, x_i^0; \theta^h) + b \quad (14)$$

As $c$ can be provided unlikely between (12) and (14), as studied in [21-22], the maximisation of likelihood is targeted via solving the following optimisation formulation ($f_3$):

$$\arg\min_{c, \zeta_u, \zeta_l} \left(\frac{1}{2}\|c\|_2^2 + \lambda \sum_i^m (\zeta_{i,u} + \zeta_{i,l})\right)$$
$$\mathbf{s.t.:} \quad y_i^0 - \tilde{f}(x) \leq \epsilon + \zeta_{i,l}$$
$$\tilde{f}(x) - y_i^0 \leq \epsilon + \zeta_{i,u} \quad (15)$$
$$\zeta_{i,u} \geq 0; \; \zeta_{i,l} \geq 0; \; i \in [1, m]$$

where $\zeta_{i,u}$ and $\zeta_{i,l}$ are slack parameters, $\epsilon$ is a tolerance margin for a smoother estimation curve, $\lambda$ shows the penalty parameter for the violation of defined constrained in (15). Algorithm 2 shows the main steps and loops of coding used for efficient sample generation (stage 2). From lines 1-8, the initial data is produced using a supervised ML (Kernel method) to adjust the samples' distance using Euclidean distance metric. From line 9-13, the non-dominated sorting genetic algorithm (NSGA-II) utilises the produced samples to find the global optimum point, where $X_k^*$ and $X_F$ are the product of stages 2. After the surrogate model is computed, the resulting data $X_F$ is re-used for unsupervised density spatial clustering (unsupervised) ML is employed for finding the pairwise distances among all the samples and their sampling density. The output of the clustering is highly affected by the distance between two samples $d_m$ which is provided through iterating over the percentile measures of the pairwise distances [23-25].

Fig. 1 presents the mechanism of the proposed MLHR system for magnet sizing purposes. The optimisation process consists of mainly three stages. In stage 1, the IPMSM is parametrically modelled, capable of simulating 2-D FEA. The equation (5) is solved subjected to its equal and unequal constraints for finding the MTPA/MTPV trajectories. Next, a DOE/R sensitivity analysis is performed to discover the magnet sizing parameters, as results shown in Fig. 3. The design variables are received as a vector $X^0$ (as training data) for the initial iteration in stage 2. In this stage, the proposed system has provided an innovative approach to improve sample efficiency, resulting in faster and more accurate optimisation process (stage 3). The initial training data $X^0$ and the related responses as $Y^0 \in \mathbb{R}^{m_0 \times n_r}$ have been computed, where $m_0$ and $n$ are the initial sample size and the number $X^0$ dimension. The surrogate models, see Eq. 12-15, have been trained for the objective function, $f_i(x)$, given in Algorithm 1, and the limit state function $g_i(x)$. Both $f_i(x)$ and $g_i(x)$ are working on the initial dataset of $D^0 = (X^0, Y^0)$. Hence, the kernel-based ML algorithm, (a supervised learning method) has received the training dataset with the dimension of $D^k = (X^k, Y^k)$. The kernel-based ML algorithm is employed to enhance the sample efficiency; therefore, the output of the block is a surrogate model which is capable of estimating $\tilde{f}_i^k$ and $\tilde{g}_i^k$ as a function of $X^0$. The $\tilde{f}_i^k$ and $\tilde{g}_i^k$ are received by the third stage of the system for sample generation. The generated sample can be used for any type of global optimisation algorithm. In this study, NSGA-II algorithm parameter settings are selected based on [22] with 100 population to find the optimum solution for the objective function given in Algorithm 2. The initial population of $N$ size is created based on $\tilde{f}_i^k$ and $\tilde{g}_i^k$, the next population is selected using non-dominated sorting to rank various frontier solutions. The mechanism is continuous for a maximum 100 of generations and ranks all the frontiers until the optimal frontier is found. The non-dominated sorting method is used at each generation to sort the new population using Pareto dominance terminology.

| **Algorithm 2**: MLHR objective function for Magnet Sizing |
|---|
| **1:** Range the variation of each input $x$: i.e. $x_i = (x_{i_{min}}, x_{i_{max}}) \in \mathbb{R}$ |
| **2:** Variable's space: $\wp = x_1 \times x_2 \times x_3 \times x_4 \times x_5 \times x_6 \times x_7 \times x_8 \subset \mathbb{R}^3$ |
| **3:** Create matrix $X^0$ with size $(nP \times jj)$; |
| **4:** Set $n$ and $n_v$ |
| **5:** *For $i = 1$ to $k$ & $j=1$ to $L$* |
| **6:** Initialise $X^0 \in \mathbb{R}^{m \times n}$ to create trading data of $D^k = \{X^0, Y^0\}$ |
| **7:** Based on $\tilde{f}_i^k$ and $\tilde{g}_i^k$ calculate Euclidean distance: $r = \|x_1 - x_2\|_2$ for each $k(x_1, x_2)$ |
| **8:** Generate samples by solving Eq. 10 |
| **9:** $\arg\min f(X, N_x, i_d, i_q) = f_1 + f_2 + f_3$ #optimisation cost function |
| **10:** If $f(X, N_x) < f(X_{opt}, N_x)$ |
| **11:** s.t. $V_{pm_{new}} < V_{pm_{old}}$ |
| **12:** $A_{\eta_{opt}} \geq A_{\eta_0}$ |
| **13:** $TPV_{opt} > \frac{T_{e_i}}{V_{pm_i}} > TPV_0$ end |
| **14:** If $X_F \in \mathbb{R}^{m_F, n} = X_{k=k_{max}}^*$ Pareto frontier, *print $X_k^*, D^{k=k_{max}}$* |
| **15:** *Elseif* compute density-based spatial clustering to find $d_m$ |
| **16:** set: $x_{k,L}^l, x_{k,L}^u$ for adoption process *end end* |
| **17:** *Else* re-simulate $f$ based on new set of $X^{k+1}$, *end* |
| **18:** Store all new sets of $X_k^*$, considering IPMSM performance |

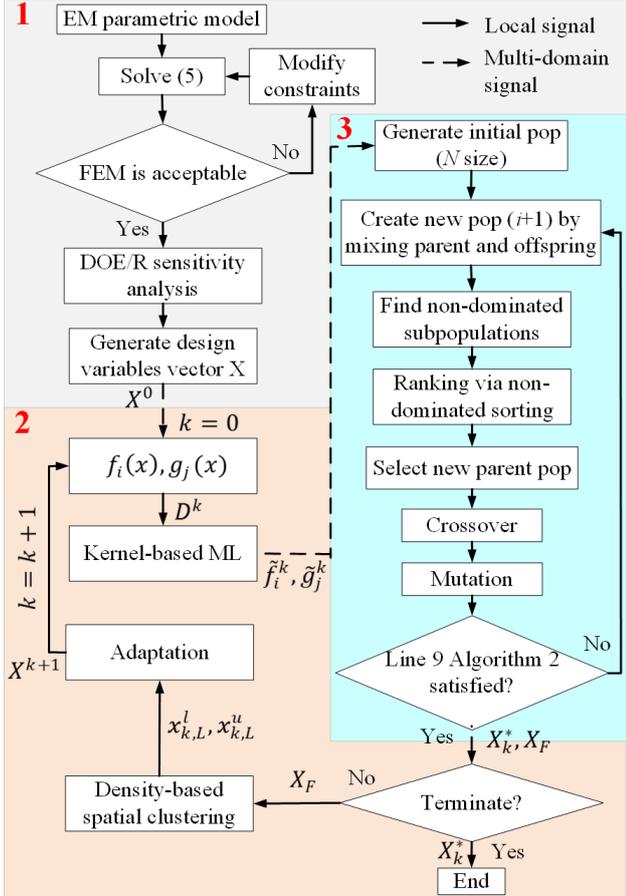

Fig. 1 The proposed MLHR system for multi-domain magnet sizing computations (1 to 3). ML: machine learning; pop: population.

## IV. RESULTS AND DISCUSSION

### A. Variable Selection and Sensitivity Analysis (Stage 1)

As shown, some of the magnet-related parameters are defined in a range for the design of experiment/ regression (DOE/R) [26-31], LHS-based sampling methods [32-36] and optimisation methods [23-25][31-33] in a pre- and post-processing stage to discover the best values for these PM-related dimensional parameters. In this study, one of the sensitivity objectives is to reduce the overall PM volume $V_{pm}$ for the modified V arrangements (Rotors 2 & 3). This consideration is performed only for the second and third rotors, not for the 3rd generation Toyota Prius reference rotor.

Fig. 2 demonstrates the studied IPMSMs with different rotor configurations, in which the second and third rotors are derived from V type magnet arrangements. The dimensions of both new second and third rotors are selected from the MLHR (as shown in Fig. 1), where the PM-focused observations $Y_i = f_i(X_i, N_{x_i})$ and trends are presented in Fig. 3. The impact of the PM thickness, $T_m$, on the maximum electromagnetic torque at the base speed is demonstrated in Fig. 3(a). Interestingly, the modified V1 (or M.V1) rotor can produce approximately the same torque with 1mm less magnet thickness, i.e. 12% less magnet volume than T-Prius (reference rotor). While M.V2 has shown a similar trend as M.V1 in terms of torque production, it can also be reported that the lowest peak-to-peak cogging torque is obtained by M.V2, with a 15mm magnet width. Both new rotors, M.V1 and M.V2, have a reduced peak-to-peak cogging torque (see Fig. 3(b)). Similarly, M.V2 has the same output power capability, using less PM width by 4mm, in contrast with the T.Prius rotor, as shown in Fig. 3(c). As the axial length of the PMs varies between 45-50mm, again M.V2 model exhibits almost the same power as the reference motor, with 3mm shorter PM axial length $L_m$. The purpose of the sensitivity is to find the influence of PM-related parameters, with a possible reduction of $V_{pm}$. In Fig. 3, the squared markings show how to obtain almost the same performance as the reference motor. Note that in this study, several simplifications have been made. For example, manufacturing tolerances, material availability, transient temperature effects, lamination material variations, radii chamfering, 3D magnetic effects, mechanical deformations and high-speed safety criteria, the influence of rotor press-fit. These final selections show a considerable magnet volume per pole reduction, in which 6.401cm³, 4.998cm³, 3.877cm³ can be reported, i.e. 12% and 40%. At least illustrating that a potential further material price reduction could be possible while maintaining the same performance factors.

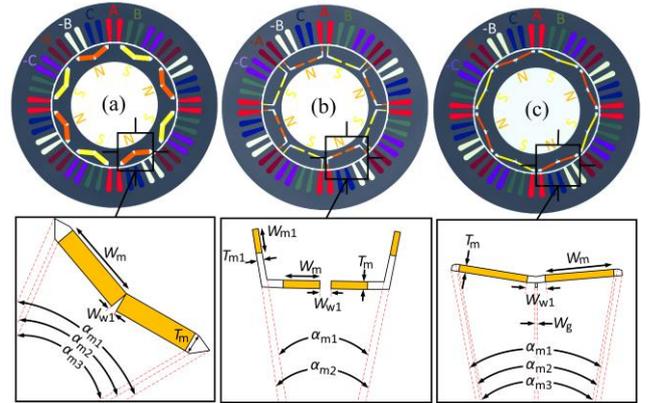

Fig. 2 PM arrangements for the studied IPMSMs, (a) 3rd Generation Toyota Prius or reference model, (b) initial modified V1, and (c) initial modified V2.

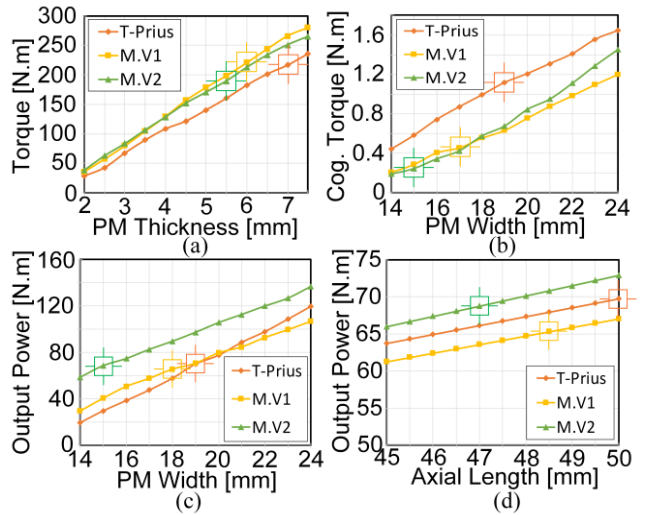

Fig. 3 Sensitivity analysis for magnet sizing, considering (a) electromagnetic torque vs. PM thickness, (b) cogging toque ripple (peak-to-peak) vs. PM width, (c) output power vs PM width, (d) output power vs. axial length. The marked squares show the selected parameter values for each configuration.

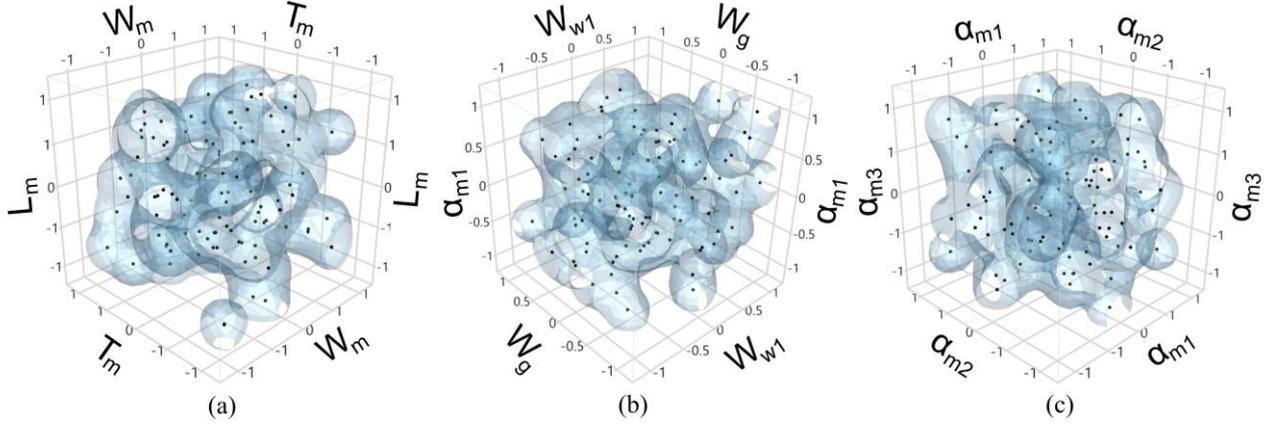

Fig. 4 Magnet design variables' sample generation using MLHR system (stage 2), focused on their distribution densities for (a) magnet width, thickness, and axial length, (b) window width, separation gap, and V shape magnet arc, (c) V shape magnet arc, V shape half flux barrier, and V shape flux barrier.

*B. Sample Generation and Optimisation Process (Stage 2-3)*

Fig. 4 presents the MLHR samples generated using Eq. 15 for the targeted controllable variables. A 95% kernel 3D contour around the samples is produced to indicate the probability density of the samples' distribution within the scales of the defined variables. It shows a darker blue colour at high density, fading to white at lower densities, and transparency that fades the contours at low density.

Fig. 5 demonstrates the optimal Pareto frontier solutions for both the LHS-based NSGA-II system and the proposed MLHR-NSGA-II system. The advantage of the proposed sample generation is demonstrated in this graph; the higher sampling efficiency has found the Pareto optimal front at a lower rank and generation number, meaning a significantly lower computation time. Both systems are converged ten times for the defined multi-objective problem, and the global optimum is found between 30-39 generations when the LHS-NSGA-II system is employed. The proposed MLHR-NSGA-II system has obtained the optimum global solution between 13-19 generations. Additionally, the number of non-dominated solutions was reported almost double that of the proposed system. The NSGA-II settings are similar for both systems. For example, the crossover and mutation probabilities are 0.8 and 0.33, the elitism rate is set to 0.55, and the maximum number of population and generation is 100. More statistical-focused details can be found in Table III. Table III summarises the best possible performances between the two systems, LHS-NSGA-II and MLHR-NSGA-II, for a better understating of the benefits gained by the proposed system. The global optimum solution is offered at the 13[th] generation using the proposed MLHR system, which is found at the 37[th] generation. The best error of 0.1 is obtained using the proposed system, and the average error is 1.95. The best and worst normalised costs (averaged) indicate that the MLHR provides better performance to reduce the cost. The sample standard deviation measurer shows a slightly better performance in LHS-NSGA-II. The function evaluation measurer reports how significantly the level of complexity is improved to 71 evaluations using the MLHR. That has improved the total computation time from 89 minutes (LHS-NSGA-II) to 40 minutes (MLHR-NSGA-II).

*C. FEA-based Results Obtained using MLHR System*

The studied IPMSMs are mapped for a better application-

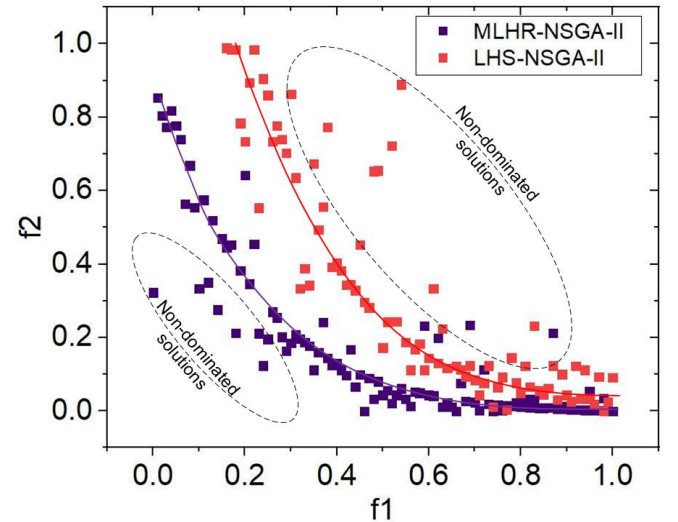

Fig. 5 Pareto optimal solutions for classic LHS-NSGA-II and the proposed MLHR-NSGA-II algorithms.

TABLE III RESULTS OF THE STUDIED SYSTEMS

| Algorithm | #Gen | Error (%) | Cost × 10^10 B | Cost × 10^10 W | $\tilde{\sigma}_h$ | $\tilde{\mu}_m$ |
|---|---|---|---|---|---|---|
| LHS-NSGA-II | 37 | 0.5-6 | 0.99 | 1 | **0.108** | 1.99×10^5 |
| MLHR-NSGA-II | **13** | **0.1-4** | **0.87** | **0.91** | 0.168 | **71*** |

*Bold numbers show the best results obtained; * indicates significant benefit.
B: normalised best cost measurer; W: normalised worst cost measurer.

oriented understanding of the optimum selection of the commutation angles. The section aims are to evaluate the electromagnetic capability, including efficiency and commutation angle maps, of both new developed IPMSM models, which ultimately allow a trade-off discussion considering the vehicle power demand.

Fig. 6 illustrates the optimum parameters offered by the MLHR for magnet and cavity sizing in 48-slot per 8-pole IPMSMs. For the T-Prius and the proposed M-V2, 12 and 14 design variables were selected, respectively. Details of the geometric parameters are given in Table IV. All simulations

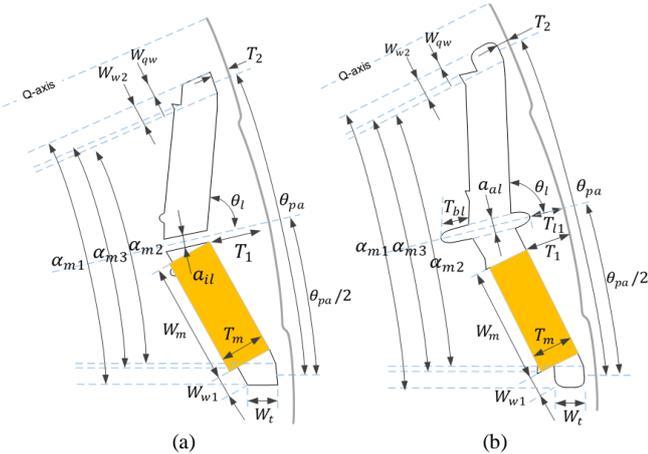

(a) (b)
Fig. 6 Schematic of the optimum magnet and its cavity parameters found using MLHR. (a) T-Prius, and (b) the proposed M-V2.

TABLE IV OPTIMUM VALUES BY MLHR FOR M-V2 MODEL

| Parameter | Description | Unit | T.Prius | M-V2 |
|---|---|---|---|---|
| $W_m$ | Magnet width | mm | 17.88 | 15.65 |
| $T_m$ | Magnet thickness | mm | 7.16 | 6.7 |
| $W_t$ | Thickness of the window | mm | 4.51 | 4.65 |
| $W_{w1}$ | Inner window width | mm | 3.8 | 3.9 |
| $W_{w2}$ | Outer window width | mm | 4 | 3.9 |
| $T_1$ | Thickness of rotor yoke under the magnet | mm | 9.56 | 8.14 |
| $T_{l1}$ | Thickness of yoke at d-axis | mm | - | 50 |
| $T_2$ | Outer width of iron ribs | mm | 2.9 | 2.8 |
| $\theta_{pa}$ | Pole arc for the whole cavity | ED | 128 | 130 |
| $\theta_l$ | Pole arc at the pole leg | MD | 72 | 73 |
| $a_{al}$ | Air gap length of the leg | mm | - | 2 |
| $a_{il}$ | Iron length of the leg | mm | 1.4 | - |
| $W_{qw}$ | Q-axis width from the PM | mm | 13.9 | 9.7 |
| $\alpha_{m1}$ | Outer cavity arc | MD | 36.15 | 38.22 |
| $\alpha_{m2}$ | Middle cavity arc | MD | 35.94 | 37.78 |
| $\alpha_{m3}$ | Inner cavity arc | MD | 36.05 | 37.89 |
| $V_m$ | Magnet volume per pole | cm^3 | 12.802 | 10.485 |
| $V_p$ | Angular pitch | Deg. | 45 | 45 |

Note: ED and MD indicate electrical and mechanical degree.

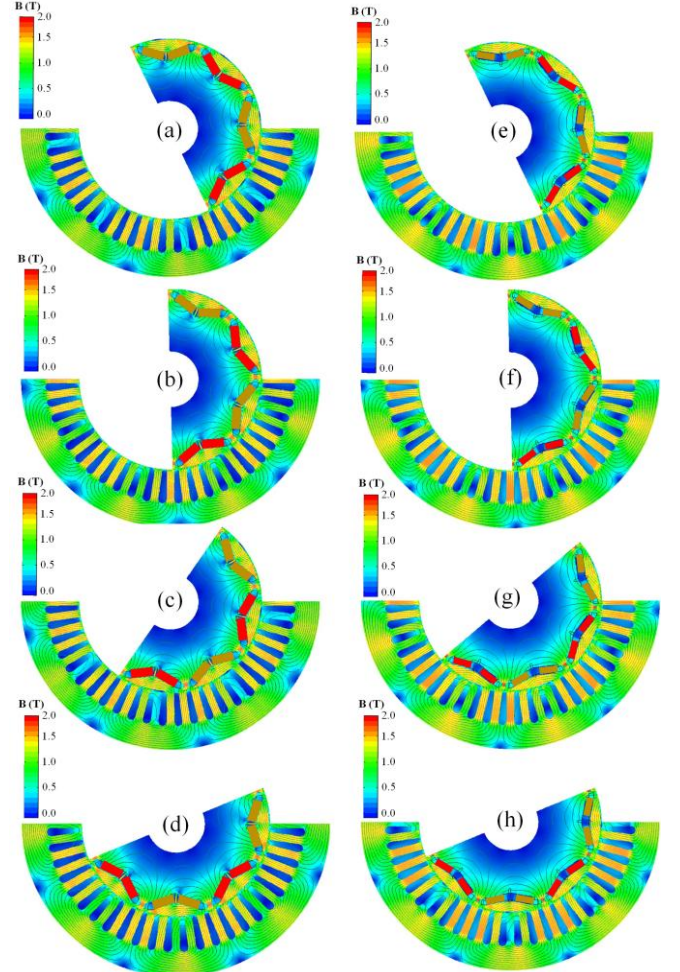

Fig. 7 Transient magnetic field computation under half load condition, between 0-180° rotor position. (a-d) T-Prius, and (e-h) the proposed M-V2.

are performed using 2-D FEA for the same materials without considering the previously mentioned variations. Also, both motoring and generating efficiencies, which can be different in practice, are assumed to be identical for simplicity. Fig. 7 demonstrates transient magnetic field distribution computed for both the reference T-Prius (see Fig. 7(a-d) and the proposed M-V2 model is presented in Fig. 7(e-h) where the rotor position varies up to 180°. The graphs confirm that the proposed M-V2 provides a slightly lower magnetic field between the magnet blocks ($a_{al}$). In addition to, the outer width of iron ribs ($T_2$) which is selected smaller considering the saturation risk. The weaker magnetic field produced using the optimum model of M-V1 has resulted in lower torque and a smaller premium efficiency region (see Fig. 9b).

Fig. 8 presents the electric vehicle (EV) configuration. In this vehicle, two onboard IPMSMs, IPMSM1 and IPMSM2 at the rear and forward wheel drives are fed via two three-phase inverters based on the MTPA/MTPV trajectory. As e-drive system, they deliver the traction force for the four-wheel-drive electric passenger car with two single-gear onboard powertrains [14,19]. Also, the vehicle utilises an electric continuously variable transmission system. Two different suspension systems are employed for the rear and forward wheels. The front wheel is a MacPherson strut with an anti-roll bar, coil springs and dampers. The rear suspension system is a Torsion beam with coil springs and dampers. The high-voltage battery unit type is Nickel-metal hydride, with a nominal voltage of 201.6Vdc and a capacity of 6.5Ah. The vehicle's maximum speed is 112mph with 0-62mph acceleration in 10.4sec. More details of the vehicle dynamics are summarised in Table V, in which most of the parameters are the same as Toyota Prius car.

Fig. 9 illustrates the efficiency maps produced by the studied IPMSMs to assess each rotor configuration's premium efficiency region ($\geq 0.94$). The focus is to obtain the greatest premium efficiency region within the constant torque and the constant power regions for all passenger vehicles. In Fig. 9(a), the efficiency map of the reference motor shows a wide premium region between starting torque (maximum 211.6Nm) and the low speed and accelerating regions. In the T-Prius motor, a PM-related torque density of 4.13Nm/cm$^3$ can be reported. The torque production has decreased to 193.5Nm, in the modified V1 motor, presented in Fig. 9(b). However, the torque density has slightly improved to 4.84Nm/cm$^3$. In this configuration, the premium efficiency region has been reduced. In the proposed rotor configuration, the M.V2 motor, with the maximum torque of 200.5Nm, and a torque density improvement of 6.46Nm/cm$^3$, has shown a better usage of PMs. Additionally, the premium efficiency region is slightly enhanced compared to the other two models (see Fig. 9(c)). The operating points are scattered for three different standard drive cycles: (i) US06, (ii) Artemis Urban, and (iii) Artemis HighWay. The US06 cycle is an aggressive, high speed (max 129.2km/h) and/or high acceleration driving behaviour, rapid speed fluctuations, and driving behaviour following start-up. Artemis cycles are produced based on European real-world driving patterns, where the Artemis_Urban and Artemis_HighWay reach 57.3km/h and 131.4km/h.

Fig. 10 presents the commutation angle maps for the studied IPMSMs, where the maps are generated to show the EV's operating points are between 0 to 212Nm with intervals of 5Nm. As a result of curve fitting a variable current amplitude, and commutation angle and corresponding control trajectory, the maps are provided for a better understanding of the $\gamma$ angle distribution within the torque-speed profile when an identical MTPA/MTPV control strategy is used for all the IPMSMs. The machine's characteristics show that the premium efficiency area is concentrated with the minimised commutation angle, which minimises the total power loss. In these graphs, the commutation angle strategy maps are presented for different driving modes: (i) urban, (ii) rural-urban, and (iii) motorway. Under the US06 cycle, most operating points have fallen within the lowest $\gamma$, indicating the premium efficiency region. Therefore, the proposed M.V2 has shown the highest premium efficiency coverage.

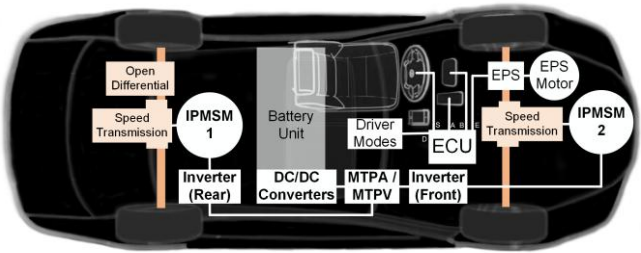

Fig. 8 Schematic of the considered EV layout.

TABLE V ELECTRIC VEHICLE SPECIFICATIONS

| Params | Description | Unit | Value |
| --- | --- | --- | --- |
| $m_0$ | Vehicle mass | kg | 1805 |
| $m_1$ | Payload | kg | 500 |
| $L_T$ | Total length | mm | 4460 |
| $h_T$ | Total height | mm | 1490 |
| $W_T$ | Total width | mm | 1745 |
| $A$ | Frontal area | m$^2$ | 2 |
| $L$ | Wheelbase | m | 2.7 |
| $a$ | Front semi-wheelbase | m | 1.3 |
| $H_{CG}$ | Centre of gravity height | m | 0.5 |
| $R_w$ | Wheel radius | m | 0.381 |
| $C_d$ | Aerodynamic drag coefficient | - | 0.25 |
| $C_r$ | Rolling resistance coefficient | - | 0.015 |
| $K$ | Rolling resistance coefficient | s$^2$/m$^2$ | 6.5e-6 |
| $SR$ | Steering ratio | - | 14.6:1 |

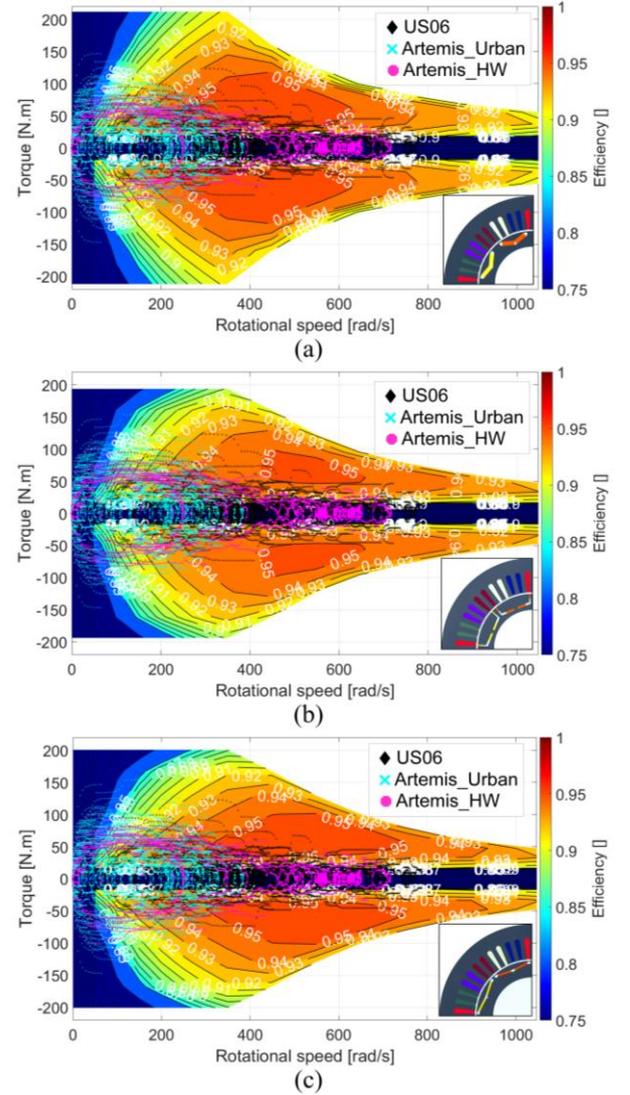

Fig. 9 Torque-speed-efficiency maps for the studied IPMSMs, (a) Toyota Prius IPMSM, (b) M.V1, and (c) M.V2; under US06, Artemis_Urban, and Artemis_HighWay cycles.

Under the Artemis Urban cycle, both Toyota Prius and M.V2 machines have performed equally well to drive the highest number of operating points within the premium efficiency. During the Artemis HighWay cycle, the M.V2 has included the most operating points at the premium efficiency region compared to the Toyota Prius machine. The proposed M.V2 model has displayed the largest premium efficiency region, particularly at high speeds, beyond the demand speeds of all the drive cycles considered, see Fig. 9(c). M.V2 model used the lowest overall commutation angle, see Fig. 10. Note that the drive switching frequency are chosen to be relatively high for automotive applications and varies between 7-8 kHz for the commutation angle maps presented. Although this will most likely result in a hearable noise from the traction electrical machine.

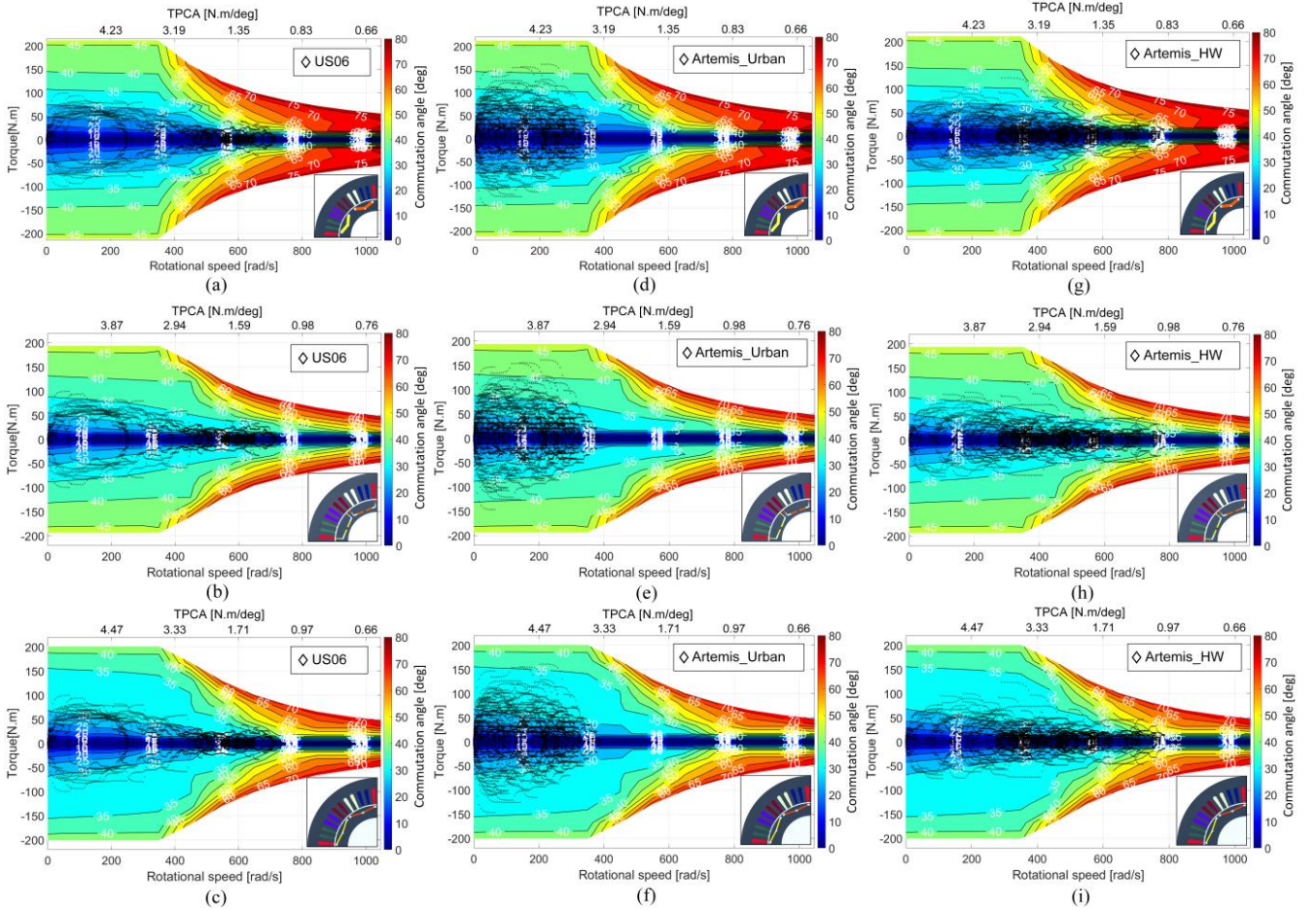

Fig. 10 Commutation angle maps for IPMSMs, considering TPCA factor, under (a) Toyota Prius or reference model under US06, (b) M.V1 under US06, (c) M.V2 under US06, (d) Toyota Prius under Artemis Urban, (e) M.V1 under Artemis Urban, (f) M.V2 under Artemis Urban, (g) Toyota Prius under Artemis HighWay, (h) M.V1 under Artemis HighWay, and (i) M.V2 under Artemis HighWay.

A new factor TPCA is defined for all three main operational regions. This ratio measures the maximum torque produced over the resulted commutation angle $\gamma$ at every specific speed demand. As the speed increases, the inverter injects a higher phase current amplitude and commutation angle to produce the required torque from the vehicle's throttle. The angle $\gamma$ varies linearly in the constant torque region; however, it rises nonlinearly during the accelerating phase. The TCPA is calculated and presented in Fig. 10 for each IPMSM configuration, and the proposed M.V2 motor achieved the highest TPCA of 4.47. During the accelerating phase (between 380-650rad/s or 3629-6207rpm), the highest TPCA of 5.04 is again obtained for the proposed M.V2 motor. Toyota Prius and M.V1 models have shown similar TPCA of approx. 4.54. During high-speed operations (between 650-1000rad/s or 6207-10000rpm), the lowest TPCA is reported for the Toyota Prius machine; the M.V1 model offers the highest TPCA. Overall, within all three regions, the proposed M.V2 has achieved the highest TPCA of 11.14 compared to the other two models.

Table VI reports the overall outcomes of the maps shown in Figs. 9 and 10. The TPCA ratio is reported for three levels of rotational speeds, low at 1000 rpm, medium at 5000 rpm, and high at 10,000 rpm. Among them, the bold rates show highest TPCA rates are obtained using M-V2 motor. The average efficiency is given for all the operating points produced by the three standard drive cycles. The highest

TABLE VI Commutation Angle Impact for Different Models

| Model | Speed | TPCA (-) | Average operating points efficiency (%) | #SiPEff (-) |
|---|---|---|---|---|
| T-Prius | Low | 4.23 | 86 | 867 |
| | Medium | 2.27 | | |
| | High | <u>0.64</u> | | |
| M-V1 | Low | <u>3.87</u> | 84 | <u>675</u> |
| | Medium | <u>2.26</u> | | |
| | High | 0.76 | | |
| M-V2 | Low | **4.47** | **88** | **877** |
| | Medium | **4.95** | | |
| | High | **0.66** | | |

Note: Best results are shown in bold, the worst are underlined.

overall efficiency is offered by M.V2 due to a higher number of samples (generated using three standard drive cycles) in the premium efficiency, SiPEff, region. The poorest performance is underlined by the M-V1.

Fig. 11 presents the FEA results for all machines studied the phase voltage under open-circuit condition (back-EMF) is shown in Fig. 11(a), where the proposed M-V2 machine obtained the peak voltage of 58.2V followed by the other innovative designs. The total harmonic distortion (THD) is 17.73%, 17.53%, and 17.12% for the T-Prius, M-V1, and M-V2 machines, respectively. Hence, the proposed M-V2 design has provided a lower THD than other solutions. The open-circuit air-gap flux density waveforms are compared in Fig. 11(b) for all three designs. The three machines have similarly obtained a maximum airgap flux density in the range of 0.99-1.1T. Compared with other machines, the proposed M-V2 has the highest amplitude of 1.1T and lowest THD of 22.54%, while the highest THDs are reported by the T-Prius and M-V1 machine with 28.74% and 33.72%, respectively. Wide constant power range capability is an indispensable requirement for IPMSMs in EV applications. Therefore, three design's power versus rotor speed curves are evaluated. To do this, the constraints of DC voltage and maximum current amplitude are limited to 600Vdc and 200Adc, respectively. Fig. 11(c) demonstrates that all three designs can achieve a wide speed range up to 10000rpm. The maximum output/shaft power of 72.54kW is reached by the M-V2 design, slightly higher than the T-Prius model. Whereas the M-V1 has obtained a lower maximum power of about 66kW (i.e about 10% less). In Fig. 11(d), the constant torque capability of each machine design against current density is compared, where the M-V2 and T-Prius are among the best and with a good agreement. The M-V2 design can provide higher TPCA and power in constant power and torque regions. The modified V designs (M-V1 and M-V2) have considered the width of the flux iron ribs ($T_2$) with the highest saturation risk. The proposed rotors are designed in such a way that adequate electromagnetic saturation can be provided in the iron ribs to minimise the magnet leakage flux. However, decreasing $T_2$ incautiously may harm the mechanical rigidity (which is investigated and reported in Table VII). The unwanted effect of saturation, which is called partial demagnetisation, presented in Fig. 11(e) in this study, is calculated for the three designs is linearized using a simplified analysis being $(1 - \frac{B_{r2}}{B_{r1}})$, where $B_{r1}$ is the magnet residual flux density prior to current loading and $B_{r2}$ is the magnet residual flux density after the load is applied. In a post processing study, we have looked in detail using non-linear characteristics whether partial demagnetization would occur at elevated temperatures in the high grades Neodymium Boron magnets. This partial demagnetization also occurs at high commutation angles besides temperature as the peak of the demagnetization field will vary along the pole-arc of the magnet, hence this has also been researched. The magnet thickness has been reduced by about 6% compared to the original design, and inter magnet leakage has been reduced with the presented optimized design, hence the authors believe that partial demagnetization will be very low and certainly does not impact the presented results. The demagnetisation ratios, presented in Fig. 11(e), are calculated based on the method studied in [40]. When little to nonpartial demagnetisation occurs, the rate is close to 0% using this method. When the magnets are fully demagnetised as a function of temperature and/or other external factors, the rate is 100%. The machines are below 20% demagnetised when the temperature reaches about 140°C, as shown in Fig. 11(e). The commutation angles of the three designs are plotted as a function of speed in Fig. 11(f). The T-Prius model requires higher angles when working at higher speeds in the constant power region, while the M-V2 design has shown better performance for MTPA/MTPV control throughout the speed range. The lowest commutation angles for the M-V2 design across the speed envelope favour this design in terms of efficiency and protection against demagnetisation for the same magnet temperature.

The mechanical analysis of the rotors is done under the highest speed at 10,000rpm, instead of the industrial rule of thumb that considers a 10% safety margin. In this analysis, the radial and centrifugal forces are calculated as a function of rotational speed for all three machines. Based on the technique used in [41], the radial force density and its harmonic contents are computed and presented in Fig. 12(a-b) for the studied machines. The electromagnetic radial force is computed for a range of 500-3750rpm under 50N.m torque demand and switching frequency of 7.5Hz. The harmonic content shown in Fig.12(b) indicates that the highest harmonic occurs at the 12$^{th}$ order. This is a simplified analysis that does not include electric motor acceleration, tolerances, material imperfections, unbalance, rotor eccentricity and rotor whirling. However, the authors still

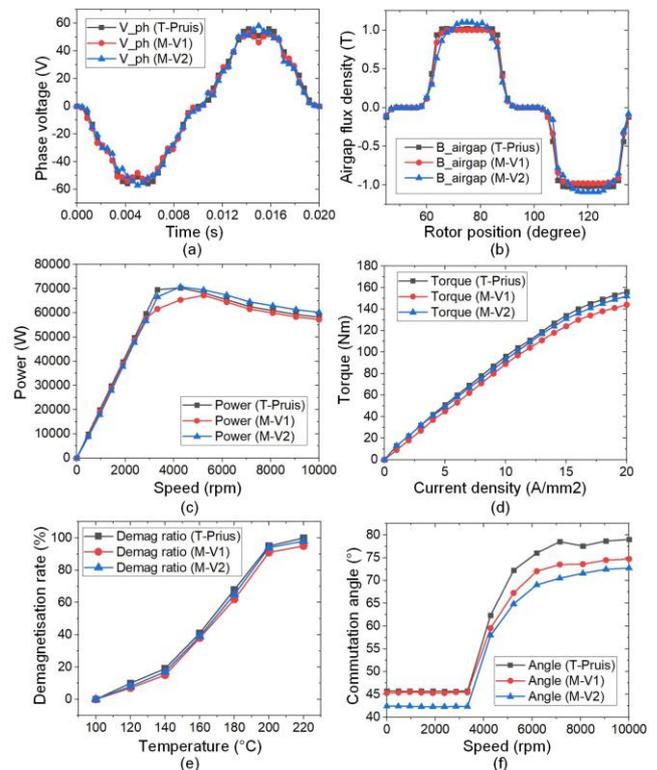

Fig. 11 Electromagnetic analysis using 2-D FEA for the studied IPMSMs; (a) phase voltage, (b) airgap flux density, (c) useful power, (d) torque as a function of current density; (e) simplified linear (partial) demagnetisation as a function of temperature, and (f) commutation angle in a wide range of operating speed.

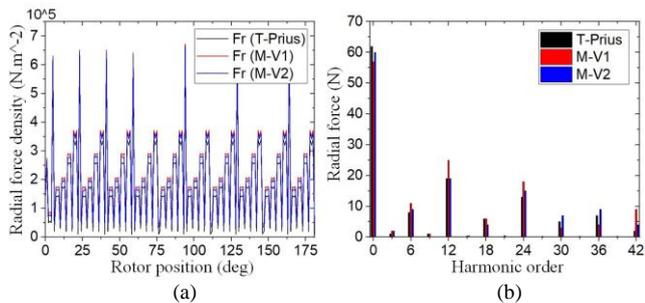

Fig. 12 Radial force analysis using 2-D FEA for the studied IPMSMs; (a) radial force density, and (b) its harmonic content.

TABLE VII ROTOR STRUCTURAL PARAMETERS AT 10000 RPM

| Model | Von Mises stress (MPa) | Displacement (mm) | Noise (dB) 0-10 kHz | Noise (dB) 10-20 kHz |
|---|---|---|---|---|
| T-Prius | 178.87 | **0.351** | **63.65** | **41.76** |
| M-V1 | <u>207.23</u> | <u>0.595</u> | <u>74.76</u> | <u>47.82</u> |
| M-V2 | **177.54** | 0.362 | 64.15 | 42.44 |

Note: Best results are shown in bold, the worst are underlined.

believe this to be a useful analysis for inclusion in this paper. Maximum von-Mises stress and displacement in rotor cores of the proposed M-V1 and the T-Prius 2010 (reference) at 10,000rpm are compared in Table VII. The table validates that maximum stress for the optimal M-V2 design is minimised compared to the other two models (M-V1 and T-Prius 2010). The highest stress has been seen at the magnet window cavities and the legs (or the upper bridges between the magnet blocks). Among all the models, the maximum stress regions are similar and reported in the iron ribs between cavities. The impact of the residual magnetism of PM material and the temperature rises when both machines are under heavy loads can be found in [39]. Additionally, the lowest displacement of 0.351mm followed by the M-V2's 0.362mm are the best values possible for the modified rotor and magnet design. The displacement produced by the centrifugal force appeared as deformation. The structural-related design is critical to consider the materials strength requirements. The radial force density contributes to the noise and vibrations in the machines, which is defined using a simplified analysis as $\sigma_{rf} = \left(\frac{B_r^2(\theta,t)}{2\mu_0}\right)$ where $B_r$ is the radial component of magnetic field density as a function of $\theta$ and $t$ are the stator reference angles. $\mu_0$ is vacuum permeability constant of $4\pi \times 10^{-7}$H/m. For simplicity, the tangential and axial components of the magnetic field density are ignored, and thus, the amplitude of $B_r$ causes the radial force density to dominate. The noise peaks are computed within two frequency bands (medium: 0-10kHz, and high:10-20kHz). Due to the rotor structure, the noise peaks occur at the medium band frequencies between 7.08 kHz and 7.88kHz; the reported frequencies are relatively similar to the switching frequency. At the higher frequency band, the noise peaks appeared at a range frequency between 14.25-15.87kHz (almost double than the switching frequency). PWM inverters with constant switching frequency generate voltages with dominant amplitudes at a few frequencies around the multiples of the switching frequency [44]. Table VII shows that the M-V2 machine is promising to obtain the lowest noise among all three machines.

Steady-state thermal analysis has been used to compare the steady-state temperature rises of both models under different torque and speeds. The structure and parameters of the cooling system of the T-Prius 2010 (reference model) are taken from [37-39] to design the thermal conductivity characteristics of materials and heat convection coefficients (see [37-38] for more details). The ambient temperature on the stator surface as an equivalence of the coolant is selected as 25 (see [39] for more information). The temperature distributions of both the reference and the proposed M-V2 machines at 150Nm and 750rpm are compared in Fig. 13.

While all the studied machines have utilised similar drive, stator geometry, and winding topology, the machine's efficiencies are suppressed by stator core, rotor core, copper, and mechanical losses. The losses are calculated using FEA and reported in Table VIII. The stator and rotor core losses are calculated using the simplified and modified Steinmetz's equation as a function of equivalent frequency (based on the remagnetisation rate of $dM/dt$ and temperature rise as presented at higher speeds. The empirical-based equation is defined for the copper loss in the stator winding, which depends on the rotor speed, stator current, and winding resistivity (AC losses are neglected). The Eddy current loss in the magnets is ignored as their values are considered to be insignificant. The mechanical loss is calculated using an empirical equation, and its value varies as a quadratic function of speed. Among all the three machines, the

TABLE VIII LOSSES IN THE STUDIED IPMSMs

| Model | Speed (rpm) | SC (kW) | RC (kW) | CL (kW) | ML (kW) | Eff (%) |
|---|---|---|---|---|---|---|
| T-Prius | 1000 | 4.975 | 0.862 | 6.855 | 1.875 | 91.21 |
|  | 5000 | 2.554 | 0.382 | 3.322 | **0.587** | 95.25 |
|  | 10000 | 2.197 | 0.341 | 1.322 | **0.549** | 92.58 |
| M-V1 | 1000 | 4.963 | 0.665 | 7.875 | 1.727 | 90.98 |
|  | 5000 | 2.546 | 0.303 | 5.324 | 0.599 | 95.06 |
|  | 10000 | 2.016 | **0.285** | 3.432 | 0.552 | 93.6 |
| M-V2 | 1000 | **4.962** | 0.664 | 6.773 | 1.726 | **91.32** |
|  | 5000 | **2.541** | 0.301 | 3.312 | 0.597 | **95.79** |
|  | 10000 | **2.003** | 0.332 | 1.318 | 0.561 | **93.85** |

Note: SC, RC, CL, and ML indicate stator core, rotor core, copper, and mechanical losses. The best achievements are indicated as bold.

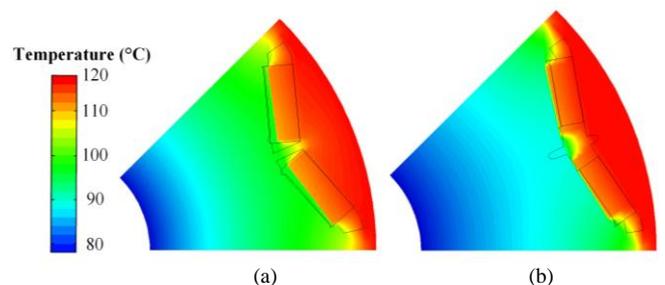

Fig. 13 Temperature distributions of the rotor using FEA resulted by the MLHR for (a) T-Prius machine and (b) the proposed M-V2 machine.

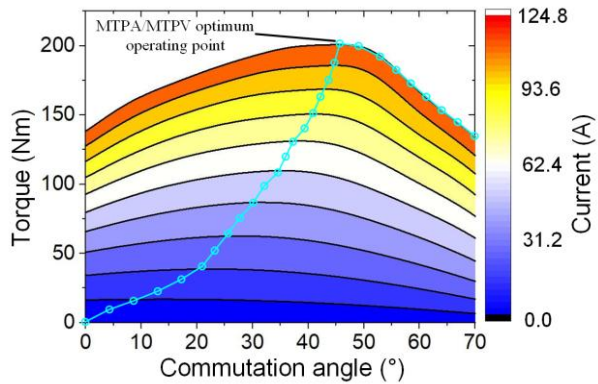

Fig. 14 Constant torque per commutation angle provided by the MLHR for the proposed M-V2 model.

proposed M-V2 design has offered a lower total loss which has reflected in a total higher efficiency in the machine. An efficiency of 91.32% (constant torque region), 95.79% (constant power region), and 93.85% (maximum speed) are reported. The results show the losses seem to be reduced within the entire wide speed range for the M-V2 design compared to the reference T-Prius machine.

Fig. 14 illustrates the MTPA/MTPV controller's actual trajectory to find the optimum operating points in the MLHR system. The system operates by obtaining the maximum torque when the machine requires less $I_d$ and $I_{ph}$. For the M-V2 design, the constant torque trajectories as a function current (with intervals of 31.2A) are presented. For the MTPA/MTPV controller, there are infinite possible choices of $V_d$, $V_q$ and $I_d$, $I_q$ to achieve the maximum torque per Ampere. The blue optimum trajectory curve indicates the selected optimum commutation angle to utilize the minimum current amplitude through a convex function $f(V_d, V_q) = I^2$ to achieve maximum torque under different loads. When the trajectory curve meets the voltage limit, the optimum trajectory can be seen within the voltage boundary.

*D. Drivability and Electromagnetic Factors Trade-offs*

In this section, the longitudinal acceleration and road gradient factors represent the drivability performance of the EV, based on [14,42-43]. Both factors are simulated considering a tire-road friction $\mu_{max} = 10$ and a zero slope. As two IPMSMs are used in front (F) and rear (R) axles, hence the torque at each IPMSM, using a backward-facing vehicle drivetrain model, is:

$$T_m = \begin{cases} \frac{1}{2} \frac{F_t R_w}{\eta_{trans} G_r}, & if\ F_t \geq 0 \\ 0, & if\ F_t < 0 \end{cases} \quad (10)$$

The maximum longitudinal acceleration $a_{x,max}$ is computed, as an optimisation problem, under the assumption that the front-to-total motor torque distribution can vary with respect to the intervention of a traction controller that prevents wheel spinning on the critical axle:

$$\begin{cases} a_{x,max} = \max_{T_{m,F}, T_{m,R}} \frac{(T_{m,F}+T_{m,R})G_r \eta_{trans} - C_r(m_0+m_1)gR_w}{(m_0+m_{app}+m_1)R_w} \\ s.t. \\ T_{m,F/R} \leq T_{m,max} \\ \frac{F_{x,F/R}}{F_{z,F/R}} \leq \mu_{max} \end{cases} \quad (11)$$

The next drivability-related factor is the maximum longitudinal road gradient, $\vartheta_{max}$:

$$\begin{cases} \vartheta_{max} = \max_{T_{m,F}, T_{m,R}} \vartheta \\ s.t. \\ \frac{(T_{m,F}+T_{m,R})G_r \eta_{trans}}{R_w} - (m_0+m_1)g\sin(\vartheta) \\ -C_r(m_0+m_1)g\cos(\vartheta) = 0 \\ T_{m,F/R} \leq T_{m,max} \\ \frac{F_{x,F/R}}{F_{z,F/R}} \leq \mu_{max} \end{cases} \quad (12)$$

where the vehicle can travel at very low speed and zero longitudinal acceleration.

Table IX reports the main parameters resulting from the drivability factors and the FEA-focused electromagnetic analysis for each IPMSM. The average machine's efficiencies are 94.1%, 93.8%, 94.9% for T-Prius, M.V1, and

TABLE IX DRIVABILITY AND ELECTROMAGNETIC MAIN FACTORS

| Parameters/ models | | T-Prius | M.V1 | M.V2 |
|---|---|---|---|---|
| Drivability and Efficiency Factors | | | | |
| $\bar{\eta}_m$ [%] | US06 | **95.0** | 94.2 | 95.0 |
| | Artemis_Urban | 95.0 | 93.8 | **95.5** |
| | Artemis_HW | 92.5 | 93.5 | **94.4** |
| $a_{x,max}$ [m/s²] | | 7.55 | 7.41 | **7.67** |
| $\vartheta_{max}$ [deg] | | 42.8 | 42.8 | **43.6** |
| Electromagnetic Factors | | | | |
| TPCA [Nm/deg] | | 10.3 | 10.1 | **11.1** |
| $T_e/V_{pm}$ [Nm/cm³] | | 4.13 | 4.84 | **6.46** |
| $T_{max}$ in PMs at base speed [°C] | | 69.7 | 67.8 | **66.9** |
| $P_{out,max}$ [kW] | | **69.6** | 59.4 | 69.3 |
| Power factor at base speed [-] | | 0.96 | 0.94 | **0.99** |
| $\beta_{max}$ at base speed [deg] | | 50.2 | 50.1 | **45.2** |

Notes: Bold text indicates the best performance among the studied IPMSMs, whereas underlined text shows the worst performance.

M.V2 models, respectively. These results demonstrate that the proposed M.V2 model has shown the best performance in drivability and electromagnetic-based investigations. It is important to mention a significant reduction in magnet size in both M.V1 and M.V2 models by 22.0% and 39.4%, respectively, compared to the 3rd generation Toyota Prius vehicle. Consequently, this leads to a substantial decrease in the machine's material cost.

## V. CONCLUSION

This study proposed a novel MLHR system for sampling improvement of the optimisation process to find the optimal sizing of V shape magnets, considering commutation angle, for the IPMSMs used in traction applications such as EVs. The main research contributions of the study are:
  i. developing MLHR system for improving sampling efficiency incorporated with the NSGA-II optimisation algorithm (e.g. magnet sizing), considering the MTPA/MTPV cost functions;
  ii. studying the impacts of a new design factor called TPCA is introduced, which derived from the commutation angle maps evaluation;
  iii. applying the studied IPMSM model at the powertrain level of a passenger EV, to map both the electromagnetic and automotive findings.

The maps and TPCA factor can help engineers in the optimal selection of drives for the electrical machines. In this work, two new IPMSM rotor configurations are proposed and compared with the 3rd generation Toyota Prius IPMSM with fixed stator configuration. The newly developed rotors based on the magnet arrangements, M.V1 and M.V2 models, are sized using a constrained LHS-based optimisation method. The efficiency and commutation angle maps were provided for the first time to understand better the operating MTPA/MTPV strategy within the wide speed range operations. The main results achieved with the proposed M.V2 model are: (i) an improvement of 36% in the torque density, while the PMs use was reduced; (ii) which will contribute to IPMSM material cost reduction; (iii) better rotor dynamic has resulted in lowering the cogging torque significantly; (iv) the only undesired finding is the maximum output power capability which has decreased by 0.43%; (v) higher electric drive efficiency at the vehicle level over the entire three drive cycles by 0.84% (in average), in which the maximum inverter current is 125A for the studied IPMSMs; (vi) enhanced maximum longitudinal acceleration at the vehicle level by 1.59%; finally (vii) the maximum longitudinal road gradient improvement by 1.85% at the vehicle level. All the IPMSM-related simulations are carried out using 2-D FEA.